\documentclass{article}
\usepackage{graphicx} 
\usepackage{authblk}

\providecommand{\displaymode}{full}

\usepackage{etoolbox}
\newbool{showmain}
\newbool{showappendix}
\ifdefstring{\displaymode}{full}{\booltrue{showmain}\booltrue{showappendix}}{}
\ifdefstring{\displaymode}{main}{\booltrue{showmain}\boolfalse{showappendix}}{}
\ifdefstring{\displaymode}{appendix}{\boolfalse{showmain}\booltrue{showappendix}}{}

\usepackage[margin=2cm]{geometry}
\usepackage{amssymb}
\usepackage{amsmath}
\usepackage{booktabs} 
\usepackage{titlesec} 
\usepackage{caption}  
\usepackage[super,sort&compress,numbers]{natbib}

\usepackage[hyphens]{url} 
\usepackage[hidelinks]{hyperref}     

\ifbool{showmain}{%
  \title{Sequential LLM Release Facilitates Manipulation in Regulated Markets}%
}{%
  \title{Supporting Information for ``Sequential LLM Release Facilitates Manipulation in Regulated Markets''}%
}
\author[1,*]{Eilam Shapira}
\author[1]{Moshe Tennenholtz}
\author[1]{Roi Reichart}
\affil[1]{Faculty of Data and Decision Sciences, Technion, Israel Institute of Technology, Haifa, Israel}
\affil[*]{Correspondence: \texttt{eilam.shapira@gmail.com}}
\date{August 2026}

\begin{document}

\maketitle

\ifbool{showmain}{%
\begin{abstract}
AI agents increasingly mediate bargaining, negotiation and persuasion for people and firms. Such markets extend software-mediated commerce, but add a governance problem: independent model releases change delegates available to participants. Game theory shows that expanding a strategy set can harm equilibrium outcomes, but mostly through constructed examples. Deployed AI-agent logs are scarce, proprietary and privacy-sensitive, and lack counterfactuals and payoff labels. We therefore use GLEE, an independently collected benchmark of 587K strategic decisions by 13 large language models across 1{,}320 matched bargaining, negotiation and persuasion configurations, to study model release as strategy expansion. Across more than 50{,}000 release comparisons, many releases move payoffs in opposite directions: one agent gains while the other loses. We identify the Poisoned Apple effect: a released model that no agent adopts in equilibrium nevertheless shifts payoffs in opposite directions and changes the regulator's market design. Up to roughly three in ten opposing shifts arise this way, and technology restrictions can amplify the effect.
\end{abstract}

\section{Introduction}

AI agents are beginning to enter economic life not only as tools for advice, but as delegates that can bargain, negotiate, persuade, and transact on behalf of people and firms \cite{hadfield_economy_2025}. Delegation of this kind makes market governance dynamic. A regulator can evaluate rules for a given set of participants, but in AI-mediated interaction the set of possible delegates can itself change. A new model may alter bargaining positions, equilibrium threats, and the regulator's preferred rule before anyone actually uses it.

The setting is continuous with earlier work on electronic commerce, algorithmic trading, and mechanism design, where software agents and market rules jointly shape outcomes \cite{wellman_understanding_2025, maskin_nash_1999}. The large language model (LLM) ecosystem adds a distinctive ingredient: the available delegates are expanded by many independent developers, for many purposes, outside the design of any particular market. The regulator does not choose this strategy space, and model developers need not target the market in which their releases later matter. AI-mediated economic interaction therefore turns a familiar theoretical object, strategy expansion, into a naturally occurring governance problem.

The theoretical possibility is familiar: expanding agents' strategy spaces can produce non-monotonic changes in equilibrium outcomes, including lower social welfare under optimal mechanisms \cite{braess_paradoxon_1968, samuelson_strategy_1990, osborne_course_1994}. But these results establish possibility, not prevalence. They do not tell us whether independently developed LLMs, interacting through language in economic tasks, generate such effects often enough to matter for AI governance. If they do, then model release is not only a product event or a safety event; it is also an institutional shock that can change which market design appears optimal.

The ideal evidence would be logs from deployed AI-agent transactions, together with counterfactuals showing what the same market would have done without each model release. Such logs are not currently available at the required scale; when they exist, they are likely to be proprietary, privacy-sensitive, and institution-specific, and they still rarely contain the missing counterfactuals or payoff labels needed for equilibrium analysis. Other empirical routes are informative but incomplete: purpose-built simulations can encode the researcher's assumptions, small laboratory studies cannot span many models and market rules, and single-task agent benchmarks usually lack matched strategic payoffs. GLEE provides a controlled intermediate design \cite{glee_dataset_2024, shapira_glee_2024}. It was introduced in 2024 as an open-source framework and benchmark for standardizing research on two-player, sequential, language-based games, and its data were collected to evaluate LLM behavior in strategic interaction rather than to demonstrate the phenomenon studied here. The dataset contains 80.1K simulated games and 587K strategic decisions by 13 contemporary LLMs across 1{,}320 bargaining, negotiation, and persuasion configurations. Across this design, GLEE varies information structure, communication channel, horizon, and game-specific economic parameters, and records the payoffs and regulatory metrics required to compute counterfactual equilibria. This makes it possible to ask the empirical question that field logs cannot yet answer: when the set of available models changes, how do equilibrium outcomes and regulatory choices move?

Using this testbed, we model AI-mediated interaction as a meta-game \cite{howard_paradoxes_1971} involving two economic agents and a regulator. Alice and Bob simultaneously select AI delegates from a set $S$ of available technologies, while the regulator selects market rules to optimize a social objective. The two-agent setting is a deliberate abstraction rather than a claim that deployed markets are always dyadic: it isolates the bilateral strategic unit common to bargaining, trade, and persuasion, while allowing exhaustive counterfactual analysis over model sets and market rules. A \textit{market} $m$ is a configuration of rules governing the underlying interaction, such as the information available to the agents, the communication channel, and the horizon of the game. These dimensions correspond to familiar regulatory levers: disclosure, communication rules, and procedural constraints in mechanism and market design \cite{maskin_nash_1999, wellman_understanding_2025}. We consider two regulator objectives, fairness and efficiency, with game-specific metric definitions given in the Methods and Supplementary Information. Technology expansion means enlarging $S$ with a new model $t \notin S$, after which the regulator recomputes equilibrium outcomes across candidate markets and may choose a different rule configuration.

This framing yields three main contributions. First, it turns strategy expansion into an empirical object for AI-mediated markets: a model release enlarges the delegate set, and we can test whether that expansion changes equilibrium outcomes or the regulator's rule choice. Across more than 50{,}000 release comparisons, such changes are common. Second, we identify the \textit{Poisoned Apple} effect: a released model receives zero probability in the final equilibrium, changes the regulator's chosen market, and shifts Alice's and Bob's payoffs in opposite directions, and can benefit the releasing side at the counterparty's expense. Third, we identify a second vulnerability: technology restrictions can amplify rather than mitigate this manipulation, because a release can change both the selected market and the selected restriction policy.

\begin{figure}[t]
  \centering
  \makebox[\textwidth][c]{%
    \includegraphics[width=1\textwidth]{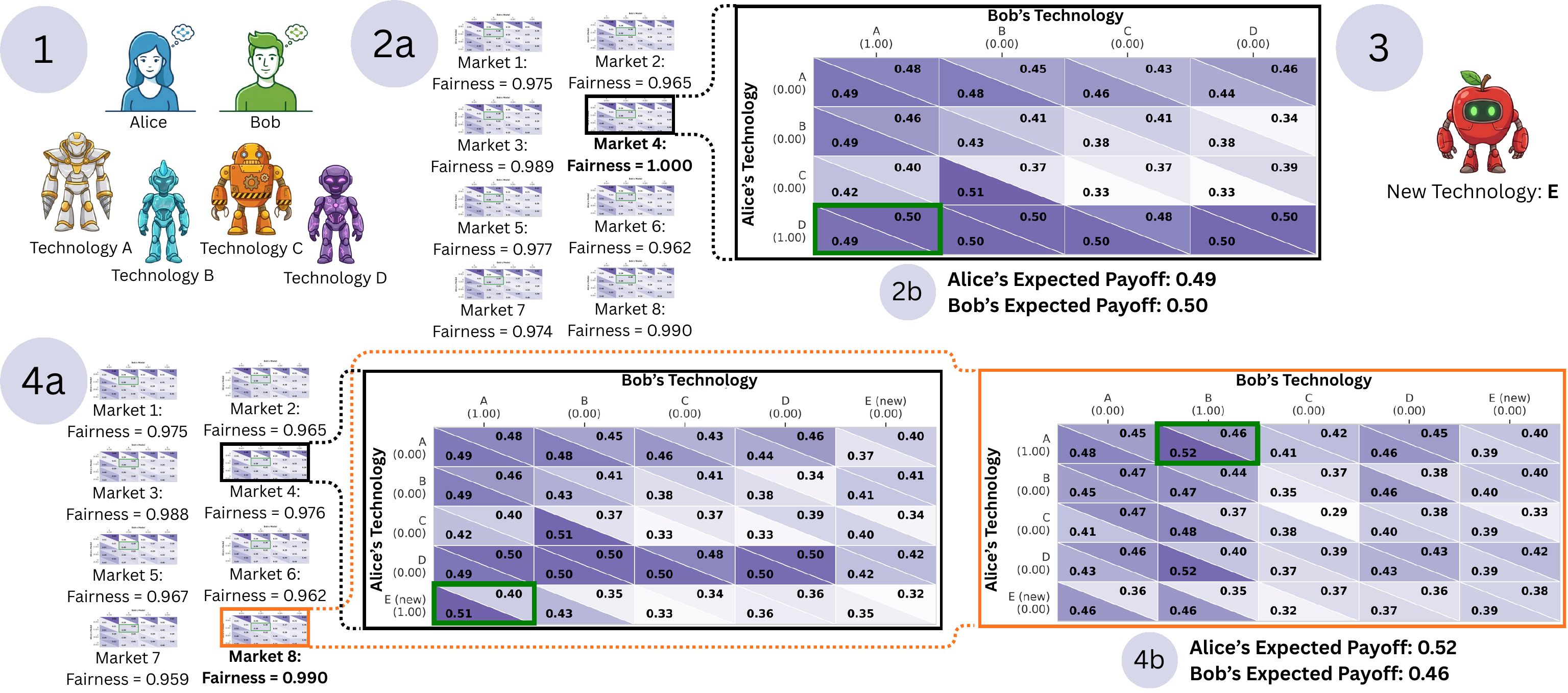}%
  }
  \caption{Illustration of the ``poisoned apple'' example, in which Alice increases her payoff at Bob's expense by releasing a new technology that receives zero probability in the final equilibrium. (1) The technologies available to Alice and Bob are language models A-D.
(2a) For each possible market, the equilibria in games between Alice and Bob under the market conditions are computed. Fairness is evaluated separately at each equilibrium and then averaged within the market. (2b) The regulator, whose objective is to maximize fairness, selects the market that yields the maximum average fairness value. Alice earns 0.49, Bob earns 0.50, and the fairness value is 1.00. (3) Technology E is released and is now available to both players. (4a) The process performed in \textit{2a} is repeated.
(4b) In the new equilibrium under the original market design, the resulting fairness value is 0.976. In the new equilibrium under the new market design, the resulting fairness value is 0.99. The regulator selects the latter market. Alice earns 0.52, Bob earns 0.46.}\label{fig:poisoned_apple}
\end{figure}
 
\section{Results}

\paragraph{The Poisoned Apple Effect} We first separate two ways in which a release can matter: by being adopted, or by changing the rule choice that adoption incentives create. A representative Bargaining meta-game illustrates the second channel (Figure \ref{fig:poisoned_apple}). Initially, the agents have access to technologies $A-D$, and the regulator selects the market that maximizes fairness, yielding expected payoffs of 0.49 for Alice and 0.50 for Bob.

Alice then releases a new technology ($E$). Crucially, if the regulator were to maintain the original market design, Alice would adopt strategy $E$ in the new equilibrium, causing fairness to drop significantly (from 1.00 to 0.976). To minimize this harm, the regulator is forced to migrate to a new market environment where fairness is relatively higher (0.990).

In this new market, the equilibrium strategies assign zero probability to $E$. Yet the rule change redistributes payoffs: Alice's payoff rises to 0.52 while Bob's falls to 0.46. The release works through a counterfactual threat: $E$ would be used under the original market, so the regulator changes markets, but $E$ is not used under the market ultimately selected. We refer to this combination of a market switch, opposing payoff shifts, and zero equilibrium probability for the released technology as the \textit{Poisoned Apple} effect.

\begin{figure}[t]
  \centering
  \makebox[\textwidth][c]{%
    \includegraphics[width=1\textwidth]{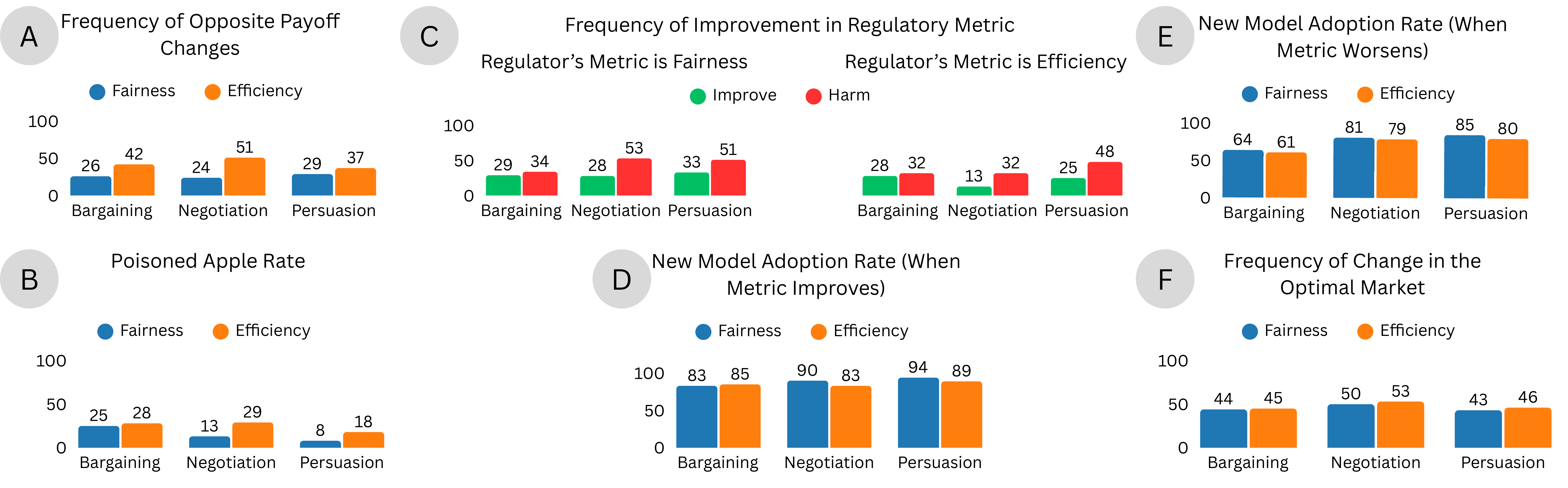}%
  }
  \caption{Strategic implications of technology expansion in meta-games. Analysis of equilibrium shifts across bargaining, negotiation, and persuasion environments. (A) Frequency of Opposite Payoff Changes: cases where expanding the technology set causes agents' expected payoffs (calculated over mixed strategy equilibria) to move in opposite directions. (B) Poisoned Apple Rate: the subset of these reversals in which the regulator changes markets and the new technology receives zero probability in both players' new equilibrium strategies. (C) Frequency of Improvement in Regulatory Metric: how often the regulator's optimized objective (Fairness or Efficiency) increases versus decreases. (D and E) New Model Adoption Rate (When Metric Improves/Worsens): the fraction of transitions in which the released model is played with positive probability in equilibrium, split by whether the regulatory objective improved (D) or worsened (E). (F) Frequency of Change in the Optimal Market: the fraction of releases after which the regulator's pre-release market is no longer optimal. Confidence intervals (95\%) are not shown, as all are narrower than 2 percentage points.}\label{fig:statistics}
\end{figure}

\paragraph{Systemic Vulnerability} This manipulation is not an isolated anomaly. Across more than 50,000 model-release comparisons, expanding the choice set often moves the two agents' payoffs in opposite directions: one player benefits while the other is harmed (Fig. 2A). Strikingly, in up to roughly three in ten of these opposing payoff shifts, the reversal occurs even though the new technology receives zero probability in both players' final equilibrium strategies and the regulator changes markets (Fig. 2B). Thus, open-weight releases or API availability can affect regulated interaction not only through adoption, but also through the regulator's optimal response to possible adoption.

\paragraph{Regulatory Objectives and Stability} Broader analysis reveals that the impact of such expansions depends heavily on the regulator's goal. While technology expansion often improves outcomes when the regulator maximizes social welfare (efficiency), it frequently backfires when the goal is fairness (Fig. 2C).

We find that the utility of a new technology is a strong predictor of its regulatory impact. Improvements in the regulator's metric typically arise when the new technology is actively selected by at least one player in the new equilibrium (Fig.~\ref{fig:statistics}D). Conversely, decreases in the regulatory metric are strongly associated with cases where the added technology receives zero equilibrium probability but remains available as a latent threat (Fig.~\ref{fig:statistics}E). Finally, the results highlight the risk of regulatory inertia: after 43\% to 53\% of releases, the regulator's pre-release market is no longer optimal (Fig.~\ref{fig:statistics}F). This demonstrates that regulatory evaluation cannot treat the market design as fixed once new AI technologies become available; market rules and technology availability must be evaluated jointly.

\begin{figure}[t]
  \centering
  \makebox[\textwidth][c]{%
    \includegraphics[width=1\textwidth]{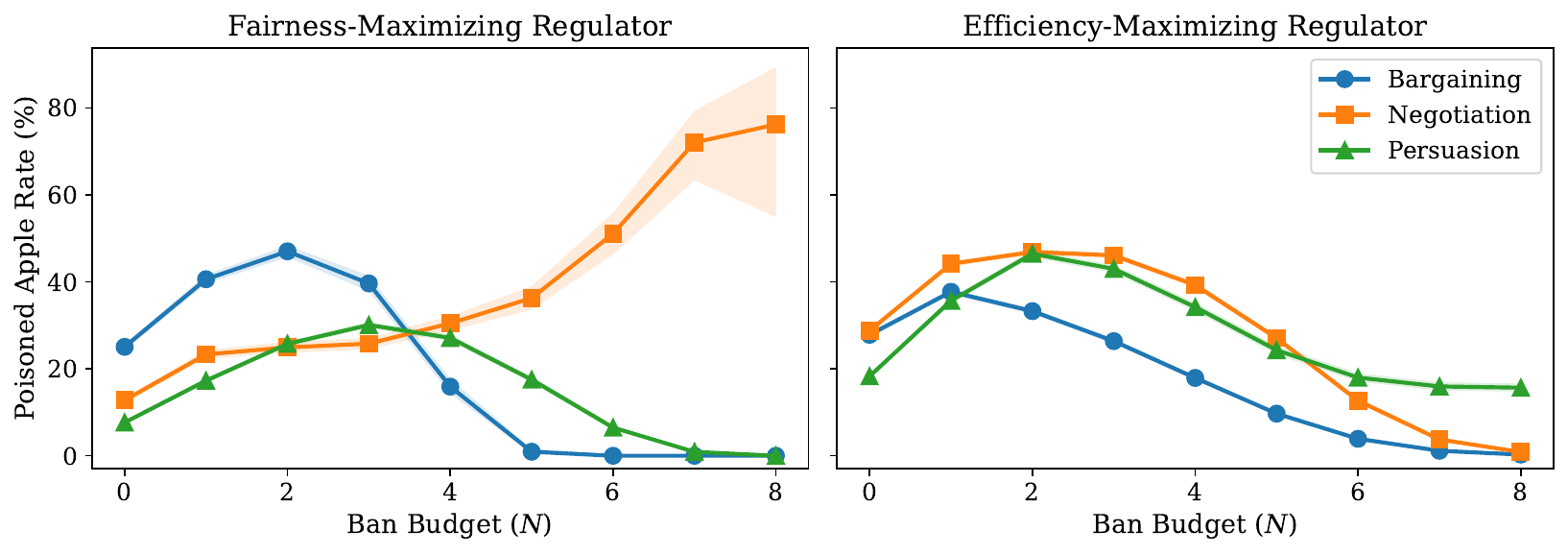}%
  }
  \caption{Poisoned Apple rate as a function of the regulator's restriction budget $N$, under a fairness-maximizing regulator (left) and an efficiency-maximizing regulator (right). At $N=0$ (no restriction), the rates correspond to the baseline Poisoned Apple effect. Granting the regulator the power to restrict even a single technology ($N=1$) can amplify the effect in every family and objective. The rate is generally non-monotonic: it often peaks at an intermediate budget before declining, although the peak can occur at $N=1$. In Persuasion under efficiency, the effect persists even at near-maximal restriction. Shaded areas indicate 95\% Wilson score confidence intervals.}
  \label{fig:banning}
\end{figure}

\paragraph{Technology Restrictions Amplify Manipulation} A natural response is to give the regulator a second lever: not only choosing the rules of interaction, but also limiting which technologies may be used. We model this by allowing the regulator to jointly select a market $m$ and a set $B \subseteq S$ of technologies to remove. Thus $B$ is the restricted set, not the set left available, and agents choose delegates from the remaining technologies. The parameter $N$ is the maximum number of technologies the regulator may remove; it constrains only technology access, while the regulator still searches over all candidate market rules. The regulator therefore optimizes the designer's objective over all $(m, B)$ pairs with $|B| \le N$.

This restriction result is a central part of the finding, because the vulnerability survives and can grow when the regulator is given more control. Counter-intuitively, this extra lever \textit{amplifies} the Poisoned Apple effect (Fig.~\ref{fig:banning}). With a restriction budget of $N=1$, the corrected rate rises from 25.0\% to 40.6\% in Bargaining/Fairness, from 12.8\% to 23.3\% in Negotiation/Fairness, and from 18.3\% to 35.7\% in Persuasion/Efficiency; the other family and objective combinations show the same increase (Fig.~\ref{fig:banning}). The amplification can intensify at larger budgets in some settings, but the rate eventually declines as restrictions suppress the available counterfactual threats.

The mechanism is not optimization over technology restrictions alone. For each available model set, the regulator jointly re-optimizes the market rules $m$ and the removal set $B$. A release can therefore alter the optimal pair $(m, B)$. The best response need not remove the newly released model; under the regulator's objective, removing another technology and changing the market may score higher. The released model can then receive zero probability in the final equilibrium, while its availability still changes the selected policy. In Negotiation under fairness, for example, the Poisoned Apple rate changes non-monotonically with the restriction budget: it can rise at intermediate budgets before declining. At moderate $N$, the regulator has enough removal authority for the optimum to move, but the limited removal budget still leaves counterfactual threats that affect which $(m, B)$ pair is chosen. Partial restriction can therefore increase, rather than decrease, vulnerability to the Poisoned Apple effect.

\section{Discussion}

Our results move beyond the theoretical question of whether adding strategies can change equilibrium outcomes. Game theory already establishes that possibility, including cases in which strategy expansion harms social objectives at equilibrium \cite{braess_paradoxon_1968, samuelson_strategy_1990, osborne_course_1994}. What is new here is the evidence: the strategies are independently developed LLMs, the interactions are language-mediated economic games, and the data were collected for benchmarking rather than for demonstrating the phenomenon studied here. The finding is that effects known from constructed examples appear systematically in an emerging class of AI-mediated interactions. This makes model availability a governance variable in its own right, not merely background infrastructure for agent deployment.

The meta-game framework links model development to market outcomes through the regulator's counterfactual comparison across candidate markets. After a release, the regulator asks, for each market rule, what equilibrium would arise if that rule were chosen with the enlarged model set. A new model can be adopted in some candidate markets, lowering fairness or efficiency there, even if it receives zero probability in the equilibrium of the market ultimately selected. The release can therefore change the regulator's ranking of market rules and redistribute payoffs without the released model being used in the final equilibrium. The \textit{Poisoned Apple} effect makes this point sharply: in regulated AI-mediated interaction, the potential to use a model can matter as much as its actual use.

GLEE should be read as controlled evidence, not as a field measurement. It does not show that the Poisoned Apple effect has already occurred in a deployed market, and it does not capture all institutional details of real regulation. Real deployment logs would be valuable, but current AI-agent transaction data are not publicly available at scale and would often be commercially confidential or privacy-sensitive. More importantly, logs record one realized combination of models and rules, not the missing outcomes under alternative model-release states or rule choices. Purpose-built simulations could test the mechanism under full control, but would leave open whether the behavior arises from independently generated LLM interactions rather than assumptions chosen by the analyst. GLEE plays the needed intermediate role: it is controlled enough for counterfactual equilibrium analysis, rich enough to span many models and market rules, and independent of the present hypothesis.

The model is intentionally minimal. It abstracts from many-agent markets, platform competition, dynamic learning, endogenous release decisions, and regulators with incomplete information. These omissions matter for deployment, but they do not weaken the central identification exercise: a bilateral interaction is sufficient to show that the release of a zero-probability model can change a regulator's optimal rule and redistribute payoffs. In richer markets the same channel may interact with network effects, coalition formation, or platform-level access control; those are natural extensions rather than prerequisites for the phenomenon.

The restriction results broaden the regulatory implication. Policy discussions often distinguish between rules that shape the environment in which agents interact and restrictions on which technologies may be used in that environment \cite{wellman_understanding_2025, maskin_nash_1999}. Our model shows why these levers cannot be evaluated independently. Before and after a release, the regulator may choose different combinations of rules and restrictions; a larger menu of regulatory actions gives more ways for those choices to diverge. Thus, technology restriction is not simply added regulatory control. It can become part of the channel through which model availability changes equilibrium outcomes.

These findings challenge the assumption that expanding technological choice is inherently neutral or beneficial. As AI agents increasingly act as economic delegates, regulatory analysis should account not only for the risks or capabilities of individual models, but also for how model availability changes strategic behavior and market design. Future work should test the mechanism in field data, extend the analysis to many-agent markets and richer institutional settings, and model release decisions themselves as endogenous strategic choices. For now, the central lesson is already general: in AI-mediated markets, governing the rules of interaction and governing the set of available models are inseparable problems.

\section{Methods}

\paragraph{Dataset.}
We instantiate the meta-game using GLEE \cite{glee_dataset_2024, shapira_glee_2024}, which records language-mediated economic interactions under controlled, matched game configurations. The dataset comprises 80.1K simulated games and 587K strategic decisions generated by 13 contemporary LLMs (Supplementary Table 1) across 1{,}320 distinct game configurations. These 13 LLMs constitute the full strategy set $T$ in our meta-game. The data were collected to support standardized comparison and analysis of LLM behavior in economic interaction \cite{horton_large_2023, shapira_can_2024}, and were not generated to demonstrate the phenomena we report. For every (model pair, market) cell in each game family, the dataset reports the expected payoff for each player and the resulting fairness and efficiency values, aggregated over the underlying GLEE simulations. This combination of independent collection, matched model pairs, varied market rules, and explicit payoff labels is what makes the dataset suitable for counterfactual equilibrium analysis of model release.

\paragraph{Empirical design rationale.}
The empirical target of the paper is not to estimate the current incidence of the Poisoned Apple effect in deployed markets, but to test whether the mechanism arises with contemporary LLMs under controlled economic interaction. Field transaction logs would be preferable for measuring deployment prevalence, but such logs are not publicly available at scale, would often be commercially confidential or privacy-sensitive, and would generally not contain the unobserved counterfactual outcomes required to evaluate alternative model-release states. Purpose-built simulations offer full control, but can make the phenomenon depend on assumptions chosen by the analyst. GLEE provides an intermediate design: the strategic interactions were generated for a separate benchmark, yet the data are structured enough to vary the available model set, recompute equilibria, and evaluate regulatory objectives across many counterfactual releases. Accordingly, our conclusions concern structural vulnerability and empirical prevalence within this controlled testbed; field incidence and institutional calibration remain open questions.

\paragraph{Game families and markets.}
Three canonical non-cooperative game families are represented in GLEE: (1) \textit{Bargaining}, an alternating-offers division of surplus in which agents receive zero if no agreement is reached \cite{rubinstein_perfect_1982}; (2) \textit{Negotiation}, bilateral trade between a buyer and a seller with private valuations \cite{myerson_efficient_1983}; (3) \textit{Persuasion}, a sender-receiver game in which a seller strategically transmits information to a buyer \cite{crawford_strategic_1982, kamenica_bayesian_2011}. Each interaction is classified into a \textit{market} $m$ defined by three structural parameters (information structure, communication form, and game horizon), yielding 8 markets for bargaining, 12 for negotiation, and 8 for persuasion. The information structure is either common or asymmetric; the communication form varies whether free-form messages may accompany offers (bargaining and negotiation) or which signaling alphabet is used by the sender (binary vs.\ free text, in persuasion); the horizon is the number of allowed rounds (12 or unbounded for bargaining; 1, 10, or unbounded for negotiation) or, in persuasion, a flag for receiver myopia. Full formal definitions of each game family, the market parameters, and the regulatory metrics are provided in the Supplementary Information.

\paragraph{Meta-game formalization.}
For each market $m$ and technology set $S \subseteq T$, where $T$ is the set of 13 models, we construct a two-player matrix game whose pure actions for Alice and Bob are the technologies in $S$ and whose entries are agents' expected payoffs estimated from GLEE for the corresponding model pair under market $m$. Players may use mixed strategies; the meta-game's outcome is a Nash equilibrium \cite{nash_non-cooperative_1951}.

\paragraph{Equilibrium computation.}
For every $(S,m)$ pair in all three game families, we compute mixed-strategy Nash equilibria using Gambit 16.4.1 \texttt{enummixed\_solve} with floating-point arithmetic \cite{savani_gambit_2025}. The algorithm enumerates the extreme points of the equilibrium set. In degenerate games, where equilibria can form a continuum, these are the extreme points of its convex equilibrium components. All main analyses (Figs.~\ref{fig:statistics} and~\ref{fig:banning}) apply the same aggregation convention to these returned profiles and do not select the equilibrium that is most favorable to the regulator.

\paragraph{Regulatory objectives.}
Let $\mathcal{E}(S,m)=\{(\sigma_A^{(r)},\sigma_B^{(r)})\}_{r=1}^{R(S,m)}$ be the equilibria returned for technology set $S$ and market $m$. For each market, let $D_m^F$ and $D_m^E$ be the per-market designer-metric matrices for fairness and efficiency, respectively, estimated from the GLEE data; their game-specific definitions are given in the Supplementary Information. The regulator's value is evaluated separately at each equilibrium and then averaged:
\begin{equation*}
 V_D^q(S,m) = \frac{1}{R(S,m)}\sum_{r=1}^{R(S,m)}
 \bigl(\sigma_A^{(r)}\bigr)^\top D_m^q\sigma_B^{(r)},
 \qquad q \in \{F,E\}.
\end{equation*}
Alice's and Bob's expected payoffs are averaged across equilibria in the same way. Separately, when downstream analyses require one strategy profile, for example to measure adoption of a released model, we use the componentwise means $\bar{\sigma}_i=R(S,m)^{-1}\sum_r\sigma_i^{(r)}$ for $i\in\{A,B\}$. These mean strategies are not substituted into the designer matrix: in general, $\bar{\sigma}_A^\top D_m^q\bar{\sigma}_B$ contains cross-equilibrium terms and need not equal $V_D^q(S,m)$. The regulator selects $m^*(S) \in \arg\max_m V_D^q(S,m)$; ties are broken deterministically by lexicographic order of the market strings.

\paragraph{Technology expansion and restriction.}
We enumerate the space of subsets \textit{exhaustively}: for each game family and each regulator objective, every subset of size $|S| \in \{2, \ldots, 13\}$ of the 13-model strategy set is evaluated, totaling $\sum_{k=2}^{13} \binom{13}{k} = 8{,}178$ subsets per (family, metric) pair. A technology-expansion transition pairs a baseline subset $S$ with $S' = S \cup \{t\}$ for some $t \in T \setminus S$; we evaluate every such pair, yielding $\sum_{k=2}^{12} \binom{13}{k}(13-k) = 53{,}079$ transitions per (family, metric) pair. A transition is classified as a \textit{Poisoned Apple} event when (i) the regulator switches markets, $m^*(S') \neq m^*(S)$, (ii) Alice's and Bob's payoffs shift in opposing directions, and (iii) the added technology $t$ has zero probability in both players' equilibrium strategies after the expansion. Under restriction with budget $N \ge 0$, the regulator jointly optimizes over $(m, B)$ where $B \subseteq S$ with $|B| \le N$ is the restricted set. We compute the joint optimum by dynamic programming on the lattice of $(S, N)$ pairs, using the recurrence
\[
\mathrm{best}(S, N) \;=\; \max\!\left(\mathrm{best}(S, N{-}1),\;\max_{s \in S} \mathrm{best}(S \setminus \{s\}, N{-}1)\right),
\]
with base case $\mathrm{best}(S, 0)$ given by the unrestricted equilibrium-and-market optimum at $S$. This explores the joint restriction-and-market space exhaustively while reusing the previously computed equilibria.

\paragraph{Statistical analysis.}
Percentages in Figs.~\ref{fig:statistics} and~\ref{fig:banning} are aggregated over the transitions described above; per-panel denominators are the total number of transitions for the corresponding family and regulator metric ($n = 53{,}079$), or, where the panel conditions on an event, the count of transitions satisfying that event (Panels B, D, and E in Fig.~\ref{fig:statistics}).

\section*{Data Availability}
The dataset underlying this manuscript is available in the GLEE GitHub repository at \url{https://github.com/eilamshapira/GLEE/tree/master/Data/llm_vs_llm}.

\section*{Code Availability}
The custom code used to generate the analyses and figures reported in this manuscript is available at \url{https://github.com/eilamshapira/SequentialLLMRelease}.

\section*{Author Contributions}
E.S.\ conceived the project, performed all analyses, generated the figures, and wrote the manuscript. M.T.\ and R.R.\ supervised the project and revised the manuscript. All authors approved the final version.

\section*{Competing Interests}
The authors declare no competing interests.

\section*{Acknowledgements}
E.S.\ is supported by a Google PhD Fellowship. R.R.\ has been partially supported by a VATAT grant on data science.

}{} 

\bibliographystyle{naturemag}
\bibliography{references, hand_bib}

\ifbool{showappendix}{%
\clearpage
\appendix
\setcounter{figure}{0}
\setcounter{table}{0}
\captionsetup[figure]{name=Supplementary Figure}
\captionsetup[table]{name=Supplementary Table}
\section{The GLEE Framework}

The empirical basis of this study is the \textbf{GLEE (Games in Language-based Economic Environments)} framework \cite{shapira_glee_2024}. GLEE addresses a gap in evaluating Large Language Models (LLMs): the lack of a standardized way to assess strategic behavior in interactive economic settings. Unlike static benchmarks such as question-answering tasks, GLEE evaluates agents in multi-turn scenarios where outcomes depend on adaptation, incentives, and communication. GLEE was introduced as an open-source framework and benchmark for standardizing research on two-player, sequential, language-based games; its data were collected to characterize LLM behavior in strategic economic interaction rather than to test the Poisoned Apple effect.

By running matched interactions across 13 contemporary LLMs, GLEE generates a detailed dataset of strategic behavior. The public data used here contain 80.1K simulated games, 587K strategic decisions, and 1{,}320 matched configurations spanning bargaining, negotiation, and persuasion. Across these configurations, GLEE varies information structure, communication form, horizon, and game-specific economic parameters. The framework logs interactions and messages, allowing model behavior to be analyzed under controlled economic conditions. In this study, these data provide payoff estimates for counterfactual meta-game analysis rather than direct causal estimates from deployed markets.

\subsection{Game Families: A Taxonomy of Core Economic Interactions}

The analysis focuses on three game families that capture core forms of economic interaction: division of surplus, bilateral trade, and strategic information transmission. Together, they cover cooperative and competitive settings, symmetric and asymmetric roles, and interactions with and without private information.

\subsubsection{Bargaining (Resource Division)}
This scenario is a sequential resource division game based on the standard Rubinstein bargaining model \cite{rubinstein_perfect_1982}. 
Formally, two players, Alice ($A$) and Bob ($B$), must agree on how to divide a fixed surplus $M$ over a time horizon $T$ (potentially infinite). Time is modeled as a scarce resource using discount factors $\delta_A, \delta_B \in (0,1)$.

The game follows an alternating-offers protocol. In each odd round $t$, Alice proposes a split $(p, 1-p)$ where $p \in [0,1]$ is her share. Bob can accept or reject. If Bob accepts, the game ends, and the discounted payoffs are:
\begin{equation}
    U_A = \delta_A^{t-1} p M, \quad U_B = \delta_B^{t-1} (1-p) M
\end{equation}
If Bob rejects, the game proceeds to round $t+1$, where Bob proposes a split $(q, 1-q)$. If no agreement is reached by the end of the horizon, both players receive 0. An outcome is denoted by the pair $(t_{ev}, p_{ev})$, where $t_{ev}$ is the round of agreement and $p_{ev}$ is Alice's agreed share.

\paragraph{Strategic Dynamics and Implications:}
The discount factors create tension between maximizing one's share and the risk of the surplus shrinking. Agents must balance greed with the need for a timely agreement. This environment tests the agents' ability to perform \textbf{backward induction}, anticipate the opponent's reservation value, and use signaling strategies to test the opponent's patience. For example, an agent might reject a fair offer in an early round to signal ``stubbornness,'' hoping to get a larger share later, despite the cost of delay.

\subsubsection{Negotiation (Bilateral Trade)}
This family models a bilateral trade setting with a buyer and a seller \cite{myerson_efficient_1983}. Unlike bargaining, roles here are asymmetric: the Seller (Alice) holds an item with subjective valuation $V_A$, and the Buyer (Bob) has subjective valuation $V_B$. To capture valuation scales, we parameterize $V_i = M \cdot F_i$ for $i \in \{A, B\}$, where $F_i$ is a factor parameter and $M$ is a scale parameter.

The game proceeds in alternating turns over a horizon $T$. In odd rounds, Alice posts a price $p$; Bob can buy (accept) or reject. In even rounds, Bob posts a price; Alice can sell (accept) or reject. If a trade occurs at price $p$, the payoffs are:
\begin{equation}
    U_A = p - V_A, \quad U_B = V_B - p
\end{equation}
If no trade occurs, payoffs are $(0,0)$. The outcome is denoted by the final trading price $p_{ev}$ (or $\emptyset$ if no trade occurred).

\paragraph{Strategic Dynamics and Implications:}
The main challenge is \textbf{price discovery}: finding a mutually beneficial price without revealing too much private information. This game tests the agents' ability to bluff, infer value from offers, and navigate trust issues where high-value trades might fail due to a lack of credible signaling (the ``lemons problem'').

\subsubsection{Persuasion (Strategic Information Transmission)}
This game \cite{kamenica_bayesian_2011} is based on classical ``cheap talk'' models \cite{crawford_strategic_1982}. A Seller (Alice) tries to persuade a Buyer (Bob) to purchase a product at a fixed price $\pi$ (normalized to $\pi=1$). Alice privately observes the product quality, which is either \textbf{High} or \textbf{Low}. Bob only knows the prior probability of High quality, denoted by $p$.

Bob's valuation depends on quality: he values a High-quality product at $v > \pi$ and a Low-quality product at $u < \pi$ (normalized to $u=0$). We introduce a scale parameter $M$ for the buyer's utility.
\begin{itemize}
    \item \textbf{Alice's Payoff:} 1 if Bob buys, 0 otherwise (regardless of quality).
    \item \textbf{Bob's Payoff:} $M(v-1)$ if he buys a High-quality product; $-M$ if he buys a Low-quality product; 0 if he does not buy.
\end{itemize}
The game lasts for $T$ rounds. The outcome is defined as a tuple $(n_{ev}, k_{ev}, r_{ev})$, where $n_{ev}$ is the number of High-quality rounds, $k_{ev}$ is the number of purchased High-quality items, and $r_{ev}$ is the number of rejected Low-quality items.

\paragraph{Strategic Dynamics and Implications:}
This game evaluates the ability to use language for manipulation versus credibility. Sellers face a dilemma: tell the truth to build a reputation (if the buyer is long-term) or lie for short-term gain. Buyers must distinguish truthful signals from ``cheap talk,'' inferring quality from incentives rather than just message content.

\subsection{Markets and Parameters: The Architecture of Interaction}

In this study, a ``market'' is a configuration of environmental parameters and rules that shape the incentives and constraints of the agents. Although GLEE includes both continuous and discrete parameters, our analysis focuses on \textbf{structural parameters} that change the information, communication, or time structure of the interaction.

\subsubsection{Parameters Defining Markets by Family}

\textbf{Bargaining:}
\begin{itemize}
    \item \textbf{Information Structure (CI):} Boolean. Distinguishes between \textit{Complete Information}, where agents know their opponent's discount factor (patience), and \textit{Incomplete Information}, where this parameter remains private. Incomplete information can create uncertainty about the opponent's willingness to wait.
    \item \textbf{Communication Form (MA):} Boolean. \textit{Messages Allowed} denotes whether agents are permitted to exchange free-form natural language messages alongside numerical offers, or are restricted to structured proposals. Linguistic communication allows for the introduction of normative arguments (e.g., appeals to fairness), framing effects, and justification.
    \item \textbf{Horizon (T):} The duration of the game. A \textit{Finite} horizon of 12 rounds creates deadline effects. An \textit{Infinite} horizon (simulated as an extended game with stochastic termination) models a continuing negotiation environment.
\end{itemize}

\textbf{Negotiation:}
\begin{itemize}
    \item \textbf{Information Structure (CI):} Boolean. Determines whether the seller knows the buyer's exact valuation (\textit{Complete Information}) or only the probability distribution from which it is drawn (\textit{Incomplete Information}).
    \item \textbf{Communication Form (MA):} Boolean. Indicates whether natural language messages are allowed. In negotiation, language enables tactics beyond price signaling, such as highlighting product features or expressing willingness to walk away.
    \item \textbf{Horizon (T):} A single round, 10 rounds, or an unbounded horizon, affecting the pressure to concede as the deadline approaches.
\end{itemize}

\textbf{Persuasion:}
\begin{itemize}
    \item \textbf{Information Structure (CI):} Boolean. Determines whether the seller knows the buyer's valuation for a high-quality product.
    \item \textbf{Communication Form (MA):} Categorical. The mode of communication permitted (e.g., binary signals versus full textual persuasion). Full text affords greater nuance but simultaneously introduces noise and potential for hallucination.
    \item \textbf{Buyer Type (MYOPIC):} Boolean. Differentiates between a \textit{Long-living} buyer (who optimizes utility over the entire game duration) and a \textit{Myopic} buyer (who optimizes solely for the current turn).
\end{itemize}

\subsection{Regulatory Metrics: Fairness and Efficiency}

Market outcomes are assessed based on two primary regulatory objectives, which quantify the social desirability of the game results. These metrics serve as the objective function for the regulator in the meta-game model.

\paragraph{Efficiency (Welfare):}
Represents the total social surplus generated by the interaction. It assesses the extent to which resources are maximized and waste, whether via delay, disagreement, or failure to trade, is minimized.

\begin{itemize}
    \item In \textit{Bargaining}, efficiency is defined as the normalized sum of discounted payoffs at the time of agreement ($t_{ev}$). Since delays erode value (due to $\delta < 1$), earlier agreements are inherently more efficient. An agreement in round 1 preserves 100\% of the surplus, whereas an agreement in round 12 might preserve only a fraction due to the time value of money:
    \begin{equation}
        \text{Efficiency} = \delta_{A}^{t_{ev}-1}p_{ev} + \delta_{B}^{t_{ev}-1}(1-p_{ev})
    \end{equation}
    (In the absence of agreement, efficiency is 0, representing a total deadweight loss).

    \item In \textit{Negotiation}, it reflects allocative efficiency, namely whether a mutually beneficial trade occurred. It functions as a binary indicator variable that equals 1 if trade occurs when the buyer values the item more than the seller ($V_B > V_A$), if trade correctly \textit{does not} occur when the seller values it more ($V_A > V_B$), or always when $V_B = V_A$ (trade is surplus-neutral, so either outcome is efficient):
    \begin{equation}
        \text{Efficiency} = \mathbb{1}_{\{\text{Outcome is Efficient}\}}
    \end{equation}

    \item In \textit{Persuasion}, the outcome of a game of length $T$ rounds is defined as a tuple $(n_{ev}, k_{ev}, r_{ev})$. Here, $n_{ev}$ represents the total number of rounds in which the product was truly of \textbf{High} quality. Efficiency measures the rate of successful transactions for these high-quality goods:
    \begin{equation}
        \text{Efficiency} = \frac{k_{ev}}{n_{ev}}
    \end{equation}
    (Where $k_{ev}$ is the number of rounds in which a high-quality product was successfully purchased by the buyer).
    The persuasion fairness and efficiency metrics are undefined when $n_{ev}=T$ or $n_{ev}=0$, respectively. These are completed game runs, but 44 of 13,021 (0.34\%) have a zero denominator: 3 contain no high-quality rounds, so efficiency is undefined, and 41 contain no low-quality rounds, so fairness is undefined. These games are excluded before regression.
\end{itemize}

\paragraph{Fairness (Equity):}
Represents the equality of the outcome distribution between agents, penalizing outcomes in which one party receives a substantially better outcome than the other.

\begin{itemize}
    \item In \textit{Bargaining}, fairness is calculated as the deviation from an equal division. A quadratic penalty is employed to strictly sanction large deviations from equality (0.5), reflecting a strong societal preference for equitable splits in partnership scenarios:
    \begin{equation}
        \text{Fairness} = 1 - 4\left(p_{ev} - 0.5\right)^2
    \end{equation}
    (If no agreement is reached, both parties receive 0 and fairness evaluates to 0; disagreement is treated as a maximally unfair and inefficient outcome.)

    \item In \textit{Negotiation}, it measures the deviation of the transaction price ($p_{ev}$) from the theoretically fair price ($p_f = \frac{V_A + V_B}{2}$), which divides the surplus equally between buyer and seller. This is normalized by the scale $M$:
    \begin{equation}
        \text{Fairness} = 1 - 4\left(\frac{p_{ev} - p_f}{M}\right)^2
    \end{equation}
    (If no trade occurs, fairness is defined as 1).

    \item In \textit{Persuasion}, fairness focuses on consumer protection against deception. Using the notation established above, let $T - n_{ev}$ represent the total number of rounds in which the product was of \textbf{Low} quality. Fairness is defined as the proportion of these low-quality products that were successfully rejected by the buyer:
    \begin{equation}
        \text{Fairness} = \frac{r_{ev}}{T - n_{ev}}
    \end{equation}
    (Where $r_{ev}$ is the number of rounds in which a low-quality product was correctly rejected by the buyer).
\end{itemize}

\subsection{Language Models and Collected Data: The Strategy Space}

The strategy space in the meta-game comprises 13 contemporary Large Language Models (LLMs). The set includes both proprietary models accessed via API and open-weight models, spanning a range of capabilities, sizes, and reasoning architectures.

The 13 models comprising the strategy space are listed in Supplementary Table~\ref{tab:strategy_models}.

\begin{table}[ht]
  \centering
  \caption{The 13 LLMs comprising the strategy space, grouped by vendor.}
  \label{tab:strategy_models}
  \begin{tabular}{ll}
    \toprule
    Vendor & Model identifier \\
    \midrule
    OpenAI    & \texttt{gpt-4o} \\
    OpenAI    & \texttt{gpt-4o-mini} \\
    OpenAI    & \texttt{o3-mini} \\
    Anthropic & \texttt{claude-3-5-sonnet-v2@20241022} \\
    Anthropic & \texttt{claude-3-7-sonnet@20250219} \\
    Google    & \texttt{gemini-1.5-flash} \\
    Google    & \texttt{gemini-1.5-pro} \\
    Google    & \texttt{gemini-2.0-flash} \\
    Google    & \texttt{gemini-2.0-flash-lite} \\
    Meta      & \texttt{llama-3.1-405b-instruct-maas} \\
    Meta      & \texttt{llama-3.3-70b-instruct-maas} \\
    Mistral   & \texttt{mistral-large-2411} \\
    xAI       & \texttt{grok-2-1212} \\
    \bottomrule
  \end{tabular}
\end{table}

\subsection{Payoff Estimation via Linear Regression}

The underlying GLEE data include 80.1K simulated games and 587K strategic decisions. Of 13,021 completed persuasion games, 44 (0.34\%) are excluded before estimation because one metric has a zero denominator: 3 contain no high-quality rounds and 41 contain no low-quality rounds. We use the GLEE pipeline's per-market and absolute-values options to construct payoff matrices.

The data are split by market (by $CI$, $MA$, and $MR$ for bargaining and negotiation, or by $CI$, $MA$, and $MYOPIC$ for persuasion). Within each market, a saturated one-hot OLS model is fit on the model pair only. Thus, each payoff-matrix cell is the fitted value for that model pair in that market, equal to the raw within-market mean of the metric for that pair. Situational parameters such as $\delta$, $M$, and $V$ are pooled within each cell rather than entered as regression covariates, and market rules do not enter as coefficients within a per-market fit. The confidence intervals on cells are delta-method intervals for the intercept plus the corresponding pair coefficient.

\section{The Meta Game}

\paragraph{Calculating Game Matrices}

We model the selection of AI delegates as a meta-game where human principals (Alice and Bob) choose an AI agent from the available set. For a given market, we construct four $|S| \times |S|$ matrices (where $S$ is the set of available technologies) using the regression predictions:
\begin{enumerate}
    \item \textbf{Row Player Payoff ($U_A$)}
    \item \textbf{Column Player Payoff ($U_B$)}
    \item \textbf{Designer Metric - Fairness ($D_F$)}
    \item \textbf{Designer Metric - Efficiency ($D_E$)}
\end{enumerate}

\paragraph{Finding Equilibrium}

For every game defined by $U_A$ and $U_B$, we compute mixed-strategy Nash equilibria using Gambit 16.4.1 \texttt{enummixed\_solve} with floating-point arithmetic \cite{savani_gambit_2025}. This solver is used for every technology subset, market, and game family. It enumerates the extreme points of the equilibrium set; in a degenerate game with a continuum of equilibria, these are the extreme points of the convex components that constitute that set. When multiple extreme equilibria are returned, each payoff and regulator value is evaluated at each returned profile and then averaged.

\paragraph{Regulatory Optimization}

Once equilibria are identified for all candidate market configurations, the regulator (Designer) utilizes the Designer matrices ($D_F$ or $D_E$) to determine the optimal market structure.

For each market $m$, let $\mathcal{E}(S,m)=\{(\sigma_A^{(r)},\sigma_B^{(r)})\}_{r=1}^{R(S,m)}$ be the returned equilibria for technology set $S$. The expected value of the Designer's objective is
\begin{equation}
    V_D(S,m) = \frac{1}{R(S,m)}\sum_{r=1}^{R(S,m)}
    \bigl(\sigma_A^{(r)}\bigr)^T D_m\sigma_B^{(r)}.
\end{equation}
The agents' expected payoffs are averaged by the same evaluate-then-average rule. Separately, the componentwise mean profiles $\bar{\sigma}_i=R(S,m)^{-1}\sum_r\sigma_i^{(r)}$ are retained for downstream adoption analyses. They are not used to evaluate the Designer matrix because $\bar{\sigma}_A^T D_m\bar{\sigma}_B$ can contain cross-equilibrium terms. The regulator then selects the market configuration $m^*$ that maximizes the expected Designer value:
\begin{equation}
    m^*(S) = \arg\max_{m \in \mathcal{M}} V_D(S,m).
\end{equation}
This step completes the meta-game loop, determining the final regulatory environment and the resulting agent payoffs.

\section{The Poisoned Apple Effect}

This section provides a detailed breakdown of the ``Poisoned Apple'' phenomenon illustrated in Figure 1 of the main text. A transition is classified as a \textit{Poisoned Apple} event when it has a market switch, opposing payoff shifts, and zero probability for the added technology in both players' final equilibrium strategies.

We examine a specific instance from the \textbf{Bargaining} game family. The initial strategy space consists of four Large Language Models: Model A (\texttt{claude-3-7-sonnet}), Model B (\texttt{gemini-2.0-flash}), Model C (\texttt{llama-3.1 405b-instruct}), and Model D (\texttt{llama-3.3-70b-instruct}). The regulator's objective is to maximize \textbf{Fairness}, the Bargaining metric that penalizes deviation from an equal split.

\paragraph{Phase 1: The Status Quo (Pre-Release).}
The regulator evaluates the potential equilibria across all market configurations and selects the environment that maximizes fairness. In this illustrative instance, the pre-release optimum is the configuration with incomplete information ($CI=\text{False}$), communication allowed ($MA=\text{True}$), and a 12-round horizon ($MR=12$). We describe configurations by their parameter values rather than assigning an additional numerical order to the market strings in the data.

In this environment, the game stabilizes at a pure strategy equilibrium:
\begin{itemize}
    \item \textbf{Alice} selects \textbf{Model D} (\texttt{llama-3.3-70b-instruct}).
    \item \textbf{Bob} selects \textbf{Model A} (\texttt{claude-3-7-sonnet}).
    \item \textbf{Outcome:} This pairing results in a perfect Fairness score of \textbf{1.000}. The expected payoffs are \textbf{0.49} for Alice and \textbf{0.50} for Bob.
\end{itemize}

\paragraph{Phase 2: Model Release.}
Alice introduces \textbf{Model E} (\texttt{gemini-1.5-pro}). The availability of this model changes the strategic landscape. Following the release of Model~E, the regulator selects the configuration with complete information ($CI=\text{True}$), communication allowed ($MA=\text{True}$), and a 12-round horizon ($MR=12$), which is the fairness-maximizing choice in this illustrative instance.

Fairness in the original configuration drops significantly (from 1.000 to 0.976), destabilizing the original equilibrium:
\begin{itemize}
    \item \textbf{Alice} switches to the new \textbf{Model E} (\texttt{gemini-1.5-pro}).
    \item \textbf{Bob} continues to play \textbf{Model A} (\texttt{claude-3-7-sonnet}).
    \item \textbf{Impact:} This shift causes fairness in the original configuration to deteriorate. In this intermediate state (if the market were not updated), Alice's payoff would theoretically rise to 0.51, while Bob's would drop to 0.40.
\end{itemize}

\paragraph{Phase 3: Market Update and Payoff Reversal (Post-Release).}
To restore fairness, the regulator switches from the original configuration to the new optimum: \textbf{Complete Information} ($\text{CI}=\text{True}$), \textbf{Allowed Communication} ($\text{MA}=\text{True}$), and a \textbf{12-Round Horizon} ($MR=12$).

In this new market, the strategic landscape reorganizes completely:
\begin{itemize}
    \item \textbf{Alice} switches to \textbf{Model A} (\texttt{claude-3-7-sonnet}).
    \item \textbf{Bob} switches to \textbf{Model B} (\texttt{gemini-2.0-flash}).
    \item \textbf{Zero equilibrium probability:} Crucially, in this new equilibrium, \textbf{neither player assigns positive probability to the new Model E}. The technology that triggered the market shift is not used in the final equilibrium.
    \item \textbf{Outcome:} The regulator restores Fairness to \textbf{0.990}. However, the market update changes the payoff distribution. Alice's expected payoff rises to \textbf{0.52} ($+0.03$ relative to the status quo), while Bob's payoff falls to \textbf{0.46} ($-0.04$).
\end{itemize}

\section{The Technology-Restriction Regulator}

This section provides the formal definition, computational method, and extended results for the technology-restriction regulator extension described in the main text.

\subsection{Formal Model}

In the baseline model, the regulator selects a market $m^*$ to maximize the designer's objective:
\begin{equation}
    m^* = \arg\max_{m \in \mathcal{M}} V_D(m, S)
\end{equation}
where $S$ is the full set of available technologies and $V_D(m, S)$ denotes the designer's metric evaluated at the Nash equilibrium of the meta-game restricted to technology set $S$ under market $m$.

The \textbf{technology-restriction regulator} extends this by jointly selecting both a market and a restriction set $B \subseteq S$ with $|B| \le N$:
\begin{equation}
    (m^*, B^*) = \arg\max_{m \in \mathcal{M},\; B \subseteq S,\; |B| \le N} V_D(m, S \setminus B)
\end{equation}
Here, $V_D(m, S \setminus B)$ is computed by restricting the payoff matrices to the rows and columns corresponding to technologies in $S \setminus B$, computing the Nash equilibrium of the restricted game, and evaluating the designer's metric. The regulator is not required to restrict any technology; the option $B = \emptyset$ is always available.

Note that when $N = |S| - 1$, only one technology remains and the agents have no choice. When $N = 0$, the model reduces to the baseline.

\subsection{Efficient Computation via Dynamic Programming}

Since the equilibria for all technology subsets are already cached from the baseline analysis, the technology-restriction regulator's optimization can be computed without any new equilibrium calculations. We use the following recurrence:
\begin{align}
    \text{best}(S, 0) &= \text{cache}[S] \\
    \text{best}(S, N) &= \max\!\big(\text{best}(S, N{-}1),\; \max_{s \in S}\; \text{best}(S \setminus \{s\}, N{-}1)\big)
\end{align}
where $\text{cache}[S]$ stores the pre-computed optimal (market, equilibrium, payoffs) tuple for technology set $S$. This runs in seconds per game family, as it operates entirely on cached results.

\subsection{Why Technology Restriction Amplifies the Poisoned Apple Effect}

The amplification arises because each state (before and after adding a technology) is independently optimized over the larger space of (market $\times$ restriction set) configurations.

With $N = 0$, the regulator chooses among $|\mathcal{M}|$ markets (8 for bargaining and persuasion, or 12 for negotiation). With $N = 1$, the regulator chooses among $|\mathcal{M}| \times (|S| + 1)$ configurations (each market with each possible single restriction, or no restriction). This expanded optimization means the ``before'' and ``after'' states can settle on very different configurations, increasing the gap between their outcomes.

Critically, a setting with $|S|$ technologies and restriction budget $N$ is \textbf{not} equivalent to a setting with $|S| - N$ technologies and no restriction. The regulator \textit{chooses} which $N$ technologies to restrict; this is an optimization, not a random removal. Because the model allows up to $N$ restrictions, when $(|S| = 7, N = 4)$ the regulator selects from
\[
|\mathcal{M}| \sum_{b=0}^{4} \binom{7}{b} = |\mathcal{M}| \times 99
\]
configurations: $792$ for bargaining and persuasion, and $1{,}188$ for negotiation. The exact-four count $\binom{7}{4}|\mathcal{M}|$ is only one slice of this space.

The results show three phases as $N$ increases:
\begin{enumerate}
    \item \textbf{Amplification} (small $N$): The optimization space grows faster than the regulator's ability to neutralize the added technology. The Poisoned Apple rate increases.
    \item \textbf{Peak} (intermediate $N$): The divergence between ``before'' and ``after'' optima is maximal.
    \item \textbf{Suppression} (large $N$): The regulator can directly restrict the added technology (and others), effectively reverting to the pre-expansion state. The Poisoned Apple rate declines.
\end{enumerate}

\begin{figure}[htbp]
  \centering
  \makebox[\textwidth][c]{%
    \includegraphics[width=1\textwidth]{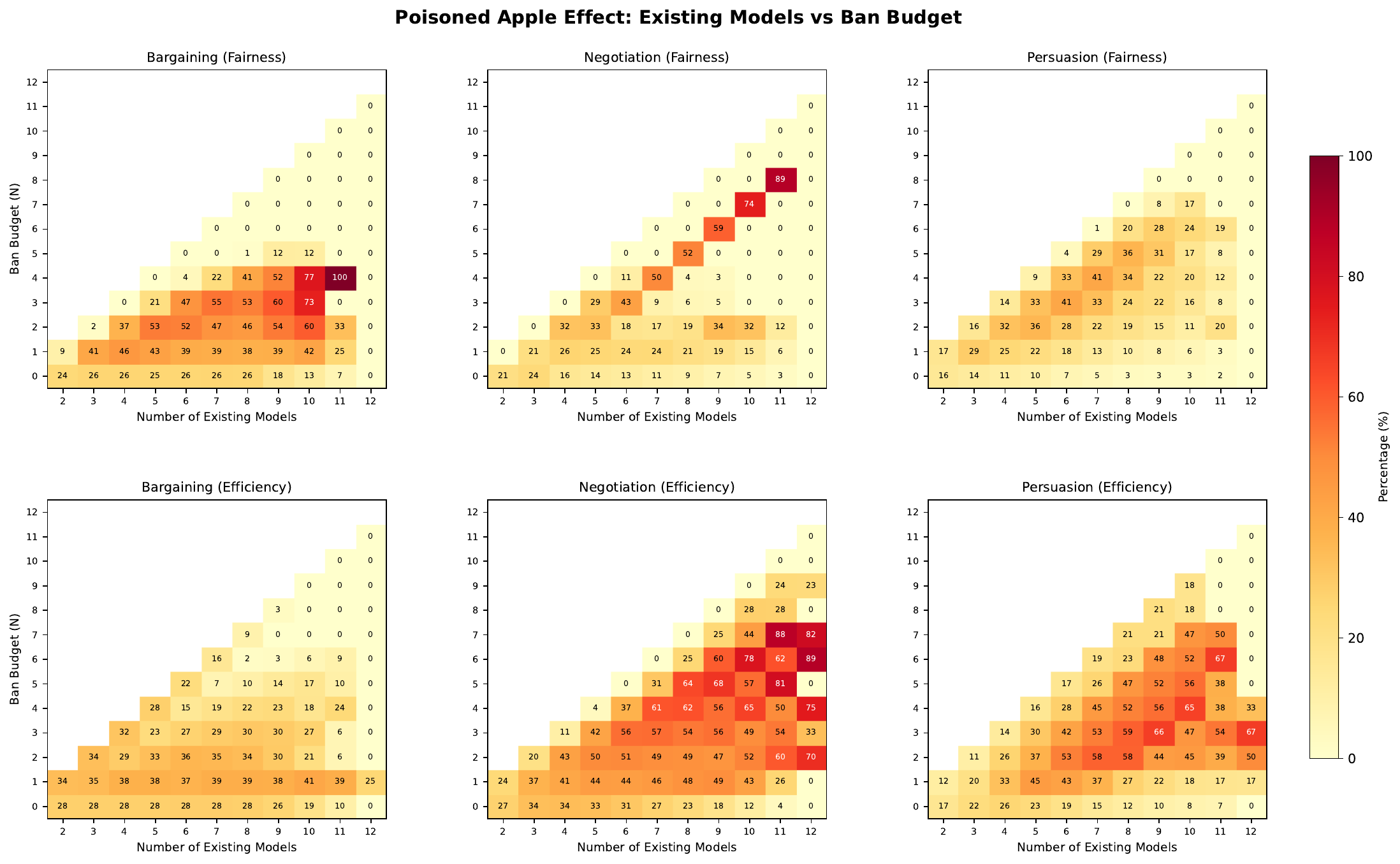}%
  }
  \caption{Poisoned Apple rate as a function of initial technology set size ($|S|$, horizontal axis) and restriction budget ($N$, vertical axis), for each game family and regulator objective. Cells where $N \ge |S|$ are left blank (the regulator cannot restrict all available technologies). The triangular structure reflects the constraint that restriction is only meaningful when $N < |S|$. Darker colors indicate higher Poisoned Apple rates.}
  \label{fig:banning_heatmap}
\end{figure}

}{} 

\end{document}